\documentclass{ws-procs9x6}

\newcommand{\cal}{\mathcal}

\newcommand{\be}{\begin{eqnarray}} 
\newcommand{\ee}{\end{eqnarray}}

\newcommand{\ssh}{/\hskip -2mm}

\newcommand{\nn}{~\nonumber \\}

\newcommand{\on}{}

\newcommand{\an}{}

\newcommand{\bmp}{\noindent\begin{minipage}{16cm}}
\newcommand{\emp}{\end{minipage}}

\newcommand{\p}{{\cal P}\exp} 

\begin{document}

\title{Fermion production in classical fields}

\author{D.~D.~DIETRICH}

\address{Laboratoire de Physique Th\'eorique, Universit\'e Paris XI, Orsay, France}

\maketitle

\abstracts{Here, the formalism for the production of fermion-antifermion pairs by vacuum 
polarisation due to classical fields is investigated according to and 
extending Refs.\cite{ddd,field}. The issue of gauge invariance is discussed for the 
calculation of the expectation value for the number of produced pairs from the exact 
retarded and/or advanced propagator. These correlators are given as solutions of the 
equation of motion for the Dirac Green's functions in the classical field. The general 
characteristics of the two point functions in field configurations expected for 
ultrarelativistic heavy-ion collisions are lined out. Calculations for a model for the 
respective radiation field are presented in a covariant-derivative expansion scheme.}

Following the derivation in\cite{derivation}, the expectation value for the number of
produced pairs can be determined by calculating the overlap of a negative energy
fermion -- corresponding to a positive energy antifermion -- with a positive energy 
fermion, after propagation through the field:
\be
M(p,q)=\int d^3x d^3y \psi^\dagger_p(x) G_R(x,y) \an \gamma^0 \on \psi_q(y).
\label{overlap}
\ee
\noindent
Use has been made of the fact that solutions of the Dirac equation at  $x$ and $y$ are 
connected by the retarded propagator which in turn can be expressed with the retarded 
single-particle scattering-operator ${{\cal T}}_R$:
\[
G_R(x,y)
=
G_R^0(x-y)
+
\int d^4\xi d^4\eta G_R^0(x-\xi){\cal T}_R(\xi,\eta) G_R^0(\eta-y).
\]
\noindent
Free solutions are linked by the free retarded propagator $G_R^0$ and positive and 
negative energy solutions are orthogonal to each other, whence integrating the square of the overlap over the phase space(s) yields:
\be
\left<n\right>
=
\int 
\frac{d^3q}{2(2\pi)^3\omega_q} 
\frac{d^3p}{2(2\pi)^3\omega_p}
\left|\bar{u}(q){{\cal T}}_R(q,-p)v(p)\right|^2,
\label{expectation}
\ee
\noindent
with the unit spinors $\bar{u}(q)$ and $v(p)$ and where $\omega_p=\sqrt{|\vec p|^2+m^2}$ and $\omega_q=\sqrt{|\vec q|^2+m^2}$. 
Gauge invariance can be discussed better in an alternative expression\cite{baltz}, 
where the derivatives act only on the two-point function:
\[
\left<n\right>
=
\int\frac{d^3q}{2(2\pi)^3\omega_q}d^4xd^4ye^{iq\cdot(x-y)}
{\rm tr}\left\{
(\ssh q+m)[i\ssh\partial_x-m]G_{+-}(x,y)[m-i\ssh\partial_y]\right\}.
\nn
\label{reduction}
\]
\noindent
Even the trace over the correlator 
\mbox{$G_{+-}(x,y)=\left<0_{in}\right|\bar{\psi}(y)\psi(x)\left|0_{in}\right>$} 
is still gauge dependent:
\mbox{$
{\rm tr}\left\{\Omega(x)G(x,y)\Omega^\dagger(y)\right\}
\neq
{\rm tr}\left\{G(x,y)\right\}
$.}

{\bf I.} One way to generate a gauge-invariant result is to project onto gauge-dependent 
solutions in a vacuum configuration of the gauge field [instead of free states as in 
Eq.(\ref{overlap})]. However, the purely partonic interpretation of the result is lost in 
the process: In all gauges other then the one in which the gauge-dependent coincides with 
the free solution, the state contains a superposition of an arbitrary number of virtual 
gluons. Further, the interpretation and the result depend on the choice of that special 
gauge. This method is equivalent to multiplying with adequate gauge (Wilson) 
links. There the choice of the path is the supplementary degree of freedom that has to be 
fixed. Additionally, in the course of the derivation of Eq.(\ref{expectation}), $q$ and 
$p$ were interpreted as the momenta of particle and antiparticle, respectively. Here, 
this stays evident only in the one gauge of reference.

{\bf II.} If purely fermionic states are to be preserved in all gauges, gauge invariance 
can be achieved by averaging the expectation value over the entire gauge group. After a 
coordinate transformation from $d^4xd^4y$ to $d^4[(x+y)/2]d^4(x-y)$ one investigates 
$d\left<n\right>/d^4(x-y)$ and finds that its average over the gauge group 
$d\left<\bar{n}\right>/d^4(x-y)$ is equal to zero but for 
$x=y$. This is similar to lattice calculations, where the final result 
contains the four-volume ${\cal V}$ of the lattice as an additional factor. Thus one 
would obtain $\left<\bar{n}\right>/{\cal V}$, there. As $x-y$ is the conjugate 
coordinate to the fermion momentum $q$ this implies that the entire dependence on that 
momentum is only due to the geometry of the phase space and the factor $\ssh q+m$ in the 
trace:
\[
\frac{d\left<\bar{n}\right>}{d^4(x-y)}
=
\int\frac{d^4q}{(2\pi)^4}\theta(+q_0)2\pi\delta(q^2-m^2){\rm tr}\{(\ssh q+m)L\}
\]
\be
L
=
\int\frac{d^4l}{(2\pi)^4}\frac{d^4k}{(2\pi)^4}
\theta(-k_0)2\pi\delta(k^2-m^2)
{\cal T}_R(l,k)(\ssh k+m){\cal T}_A(k,l)
\label{L}
\ee
The needed retarded and advanced propagators can be written as
\[
iG_{^R_A}(x,y,\vec k)\gamma^0
=
\pm\int\frac{d^3k}{(2\pi)^3}e^{i\vec k\cdot(\vec x-\vec y)}\theta[\pm(x_0-y_0)]G_H(x,y,\vec k)
\]
\noindent
with the homogeneous solution  
\[
G_H(x,y,\vec k)
= 
\p\left\{
i\int_{y_0}^{x_0}d\xi_0 \gamma^0
[i\gamma^j\partial_j(x)+\gamma^jk_j
+\ssh\hskip -0.3mm A(\xi_0,\vec x)
-m]
\right\}
\]

Now consider the special case of ultrarelativistic heavy-ion collisions. They are 
frequently described by means of colour charges moving along the light-cone\cite{kr}. 
This picture provides the adequate framework for the description of particles with low 
longitudinal momentum\cite{baltz,ddd}. The 
classical Yang-Mills equations have to be solved for that current. In Lorenz gauge the 
field consists of Weizs\"acker-Williams sheets and the radiation field inside the forward 
light-cone. The former represent the Coulomb fields of the charges boosted into the
closure of the Lorentz group, whence they are $\delta$-distributions in the light-cone 
coordinates:
\[
A_{+,-}^{WW}(x)
=
\sum_{n_{L,R}=1}^{N_{L,R}}
\delta\left[x_{-,+}-(b_{-,+}^{L,R})_{n_{L,R}}\right]
a_{n_{L,R}}^{L,R}.
\label{wwsheets}
\]
\noindent
with 
$2\pi a_{n_{L,R}}^{L,R}=-gt_a(t_a^{L,R})_{n_{L,R}}
\ln\lambda|\vec x_T-(\vec b_T^{L,R})_{n_{L,R}}|$ and where $\lambda$ regularises the 
logarithm but does not appear in the field tensor. $t_a$ are the generators of $SU(3)_c$. 
\mbox{$(t_a^{L,R})_{n_{L,R}}$} represent the colour of the 
charges. It should be 
mentioned that the colour neutrality of each nucleus has to be ensured and that the 
charges of the first nucleus precess in the field of the second and vice versa, 
causing further sheet-like contributions that can be incorporated by generalising the 
charges\cite{kr,ddd}. However, the general form of the field -- continuous
inside the forward light-cone and singular but integrable on the forward
light-cone -- allows to express the retarded propagator by a finite
number of addends\footnote{The dependence on $\vec k$ is not shown below.}:
\be
&&G_H(x,y)
=
G_H^{rad}(x,y)
+
\nn
&&+
\int d^4\xi
G_H^{rad}(x,\xi)
[T^L_\pm(\xi)\delta(\xi_-)+T^R_\pm(\xi)\delta(\xi_+)]
G_H^{rad}(\xi,y)
+
\nn
&&+
\int d^4\xi d^4\eta
G_H^{rad}(x,\xi)
T^L_\pm(\xi)\delta(\xi_-)
G_H^{rad}(\xi,\eta)
T^R_\pm(\eta)\delta(\eta_+)
G_H^{rad}(\eta,y)
+
\nn
&&+
\int d^4\xi d^4\eta
G_H^{rad}(x,\xi)
T^R_\pm(\xi)\delta(\xi_+)
G_H^{rad}(\xi,\eta)
T^L_\pm(\eta)\delta(\eta_-)
G_H^{rad}(\eta,y),
\nonumber\ee
\noindent
because higher orders in the $T$ cannot contribute, as lines of
constant $x_-$ or $x_+$ can only be crossed once\cite{baltz,ddd}. There
$G_H^{rad}(\eta,y)$ represents the homogeneous solution in the radiation field only and:
\[
T^{L,R}_\pm(x)+\rho^{+,-}
=
\rho^{+,-}
(\pm 1)^{N_{L,R}}
{\cal P}\prod_{n_{L,R}=1}^{N_{L,R}}\exp[\pm ia_{n_{L,R}}^{L,R}(\vec x_T)].
\]
\noindent
${\cal P}$ indicates that the above factors are to be
ordered in such a way that, if \mbox{$m_{L,R}>n_{L,R}$} then
\mbox{$(b_{-,+}^{L,R})_{m_{L,R}}>(b_{-,+}^{L,R})_{n_{L,R}}$}. The subscript
$\pm$ stand for the sign of the time difference \mbox{$x_0-y_0$} and 
$2\rho^\pm=1\pm\gamma^0\gamma^3$. Here, the ultrarelativistic limit has already been 
taken: In other 
words, in very energetic nuclear collisions the nuclei are highly Lorentz-contracted. For 
infinitely high energies \mbox{$\gamma\rightarrow\infty$}, the $(b_\pm^{L,R})_{n_{R,L}}$ 
go to zero like $\gamma^{-1}$. Even now, the longitudinal structure is still manifest 
in the path-ordering of the products. However, care 
has to be taken, as potentially important, because qualitatively different contributions 
to observables could  be neglected, although the expression for the propagator in this 
limit would still be correct. If finite longitudinal distances were kept, every charge 
would have to be treated on its own and the previous equation would contain contributions 
with up to \mbox{$N_L+N_R$} insertions. 

In order to proceed from here, the exact solution for the radiation field would be needed 
which requires transverse lattice calculations beyond the scope of this paper. Therefore, 
the propagator in a model for the radiation field $A^{rad}$ is to be investigated.
By virtue of longitudinally boost-invariant boundary conditions on the
light cone, it is often taken as being invariant under longitudinal boosts,
too. Although, this is not necessarily the case just such a field is chosen here:
\mbox{$\tau_0A_\pm=\pm x_\pm a\exp\{-x_+x_-/\tau_0^2\}$}, which satisfies the Lorenz, Fock-Schwinger, and $A_0=0$ gauge-conditions.

Evaluating Eq.(\ref{expectation}) for this field can be interpreted according to 
{\bf I} with the above gauge of reference. In other words, Wilson lines along the 
temporal direction have to be added to the correlator. The single-particle 
scattering-operators to the lowest order in an expansion in the transverse components of 
the covariant derivatives read
$
{\cal T}_R
=
\gamma_+{\cal T}_R^+
+
\gamma_-{\cal T}_R^-
$
where
\be
{\cal T}_R^\pm(p,q)
=
\mp a\tau_0\frac{
f(r_+r_-{\tau_0}^2)
}{
(r_+r_-{\tau_0}^2)
r_\pm
}
-a^2
\sum_{\mu\nu}
\frac{(\pm ia\tau_0)^\mu}{\mu!}
\frac{(\mp ia\tau_0)^\nu}{\nu!}
\times~~~~~~~~~~~~~~~~
\nn
\times
\frac{
f[r_+r_-{\tau_0}^2/(\mu+\nu+2)]
-
f[p_\pm r_\mp{\tau_0}^2/(\mu+\nu+2)]
}{
{r_\pm}^2
[(\mu+1)q_\mp-(\nu+1)p_\mp]
}
\nonumber
\ee
which could be partially resummed into hypergeometric functions.
$
f(\sigma)
=
\sigma^2
e^{-\sigma}
{\rm Ei}(\sigma)
-
\sigma
$
and
$r_\pm=p_\pm+q_\pm$.
Note that an expansion in covariant derivatives and/or the mass 
preserves the properties of the propagators under gauge transformations.
Evaluating the integrand yields here:
\[
\left|\bar{u}(q){{\cal T}}_R(q,-p)v(p)\right|^2
=
\sum_n[
8(q\cdot{\cal T}_R)(p\cdot{{\cal T}_R}^\dagger)
-
4(p\cdot q-m^2)({\cal T}_R\cdot{{\cal T}_R}^\dagger)
]
\]
The sum runs over the eigenvalues $\lambda_n$ of the colour matrix 
$a=\sum_n\lambda_n\left|n\right>\left<n\right|$, where all occurences of $a$ in 
${\cal T}$ are replaced by $\lambda_n$. If a light-cone gauge, {\it e.g.} $A_-=0$ was 
chosen as reference, then ${\cal T}_R^-=0$ and
\be
{\cal T}_R^+(p,q)
=
-2a\tau_0\frac{
f(r_+r_-{\tau_0}^2)
}{
(r_+r_-{\tau_0}^2)
r_+
}
-4a^2
\sum_{\mu\nu}
\frac{(+2ia\tau_0)^\mu}{\mu!}
\frac{(-2ia\tau_0)^\nu}{\nu!}
\times~~~~~~~~~~~~~~~
\nn
\times
\frac{
f[r_+r_-{\tau_0}^2/(\mu+\nu+2)]
-
f[p_+r_-{\tau_0}^2/(\mu+\nu+2)]
}{
{r_+}^2
[(\mu+1)q_--(\nu+1)p_-]
}
\nonumber
\ee
would have to be used.

The expression for the advanced single-particle scattering-operator needed in the case 
{\bf II} can be obtained 
from ${\cal T}_A(p,q)=-{\cal T}_R^*(q,p)$. For the present model, the integration over 
$d^4l$ in Eq.(\ref{L}) can be carried out analytically. However, the result -- moreso 
after carrying out the trace -- is to lengthy 
in order to be spelled out here, but will be presented elsewhere\cite{dean}. 
Further, it is not yet clear how the momentum variable $k$ in Eq.(\ref{L}) can be 
related to the antiparticle momentum $p$ in Eq.(\ref{expectation}). More reflection 
on that point is also required.

\section*{Acknowledgements}

The author would like to thank A.~Mosteghanemi for his help.
Useful and informative discussions with F.~Guerin, G.~Korchemsky, A.~Mishra, A.~Mueller, J.~Reinhardt, D.~Rischke, D.~Schiff, and S.~Schramm are acknowledged gratefully.
This project was supported financially by the German Academic Exchange Service (DAAD).


\end{document}